\documentclass[sigconf]{acmart} 
\usepackage{multirow}
\usepackage{algorithm}
\usepackage{algorithmic}
\usepackage{amsmath}
\AtBeginDocument{%
  \providecommand\BibTeX{{%
    \normalfont B\kern-0.5em{\scshape i\kern-0.25em b}\kern-0.8em\TeX}}}

\setcopyright{acmcopyright}
\copyrightyear{2018}
\acmYear{2021}
\acmDOI{XX.XXXX/XXXXXXX.XXXXXXX}

\acmConference[XXXX21]{XX}{XX}{XX}
\acmBooktitle{XXXX}
\acmPrice{XXXX}
\acmISBN{XXX-X-XXXX-XXXX-X/XX/XX}

\begin{document}

\title{AMSS-Net: Audio Manipulation on User-Specified Sources with Textual Queries}

\author{Woosung Choi}
\orcid{0000-0003-2638-2097}
\affiliation{%
  \institution{Department of Computer Science, Korea University}
   \city{Seoul}
  \country{Republic of Korea}}
\email{ws\_choi@korea.ac.kr}

\author{Minseok Kim}
\affiliation{
  \institution{Department of Computer Science, Korea University}
   \city{Seoul}
  \country{Republic of Korea}}
\email{rlaalstjr47@korea.ac.kr}

\author{Marco A. Martínez Ramírez}
\affiliation{%
  \institution{Centre for Digital Music, \\Queen Mary University of London}
  \city{London}
  \country{United Kingdom}}
\email{m.a.martinezramirez@qmul.ac.uk}

\author{Jaehwa Chung}
\affiliation{
  \institution{Department of Computer Science and Engineering, \\Korea National Open University}
   \city{Seoul}
  \country{Republic of Korea}}
\email{jaehwachung@knou.ac.kr}

\author{Soonyoung Jung}
\affiliation{
  \institution{Department of Computer Science, Korea University}
   \city{Seoul}
  \country{Republic of Korea}}
\email{jsy@korea.ac.kr}

\renewcommand{\shortauthors}{Woosung Choi, et al.}

\begin{abstract}
This paper proposes a neural network that performs audio transformations to user-specified sources (e.g., vocals) of a given audio track according to a given description while preserving other sources not mentioned in the description.
Audio Manipulation on a Specific Source (AMSS) is challenging because a sound object (i.e., a waveform sample or frequency bin) is `transparent'; it usually carries information from multiple sources, in contrast to a pixel in an image.
To address this challenging problem, we propose AMSS-Net, which extracts latent sources  and selectively manipulates them while preserving irrelevant sources.
We also propose an evaluation benchmark for several AMSS tasks, and we show that AMSS-Net outperforms baselines on several AMSS tasks via objective metrics and empirical verification.

\end{abstract}




\maketitle

\section{Introduction}

In recent days, social media applications have attracted many users to create, edit, and share their audio, audio-visual, or other types of multimedia content.
However, it is usually hard for non-experts to manipulate them, especially when they want to edit only the desired objects.
For image manipulation, fortunately, recently proposed methods such as image inpainting \cite{inpainting}, style transfer \cite{cyclegan}, and text-guided image manipulation \cite{manigan,dwtc} enables non-expert users to edit the desired objects while leaving other contents intact. 
These machine learned-based methods can reduce human labor for image editing and enable non-experts to manipulate their image without prior knowledge of tools that are often complicated to use.

On the other hand, little attention has been given to machine learning methods for automatic audio editing.
It is challenging to edit specific sound objects (e.g., decrease the volume of \textit{cicada buzzing noise}) with limited tools in the given audio.
Considering that audio editing usually requires expert knowledge of audio engineering or signal processing, we explore a deep learning approach in conjunction with textual queries to lessen audio editing difficulty.

This paper addresses Audio Manipulation on Specific Sources (AMSS), which aims to edit only desired objects that correspond to specific sources, such as vocals and drums, according to a given description while preserving the content of sources that are not mentioned in the description. 
We formally define AMSS and a structured query language for AMSS in \S \ref{sec:AMSS}.
AMSS can be used for many applications such as video creation tools making audio editing easy for non-experts.
For example, users can decrease the volume of drums by typing simple textual instructions instead of time-consuming interactions with digital audio workstations.

\begin{figure}[t]
    \centering
    \includegraphics[width=\columnwidth]
    {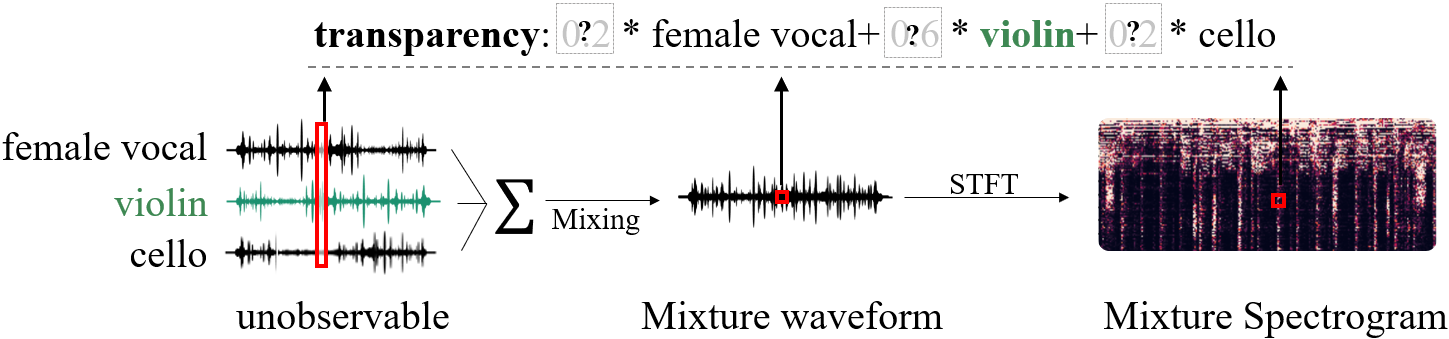}
    \caption{Sound objects are transparent}
    \label{fig:challenge}
\end{figure}

Although many machine learning approaches have been proposed for audio processing \cite{marco0,marco1,marco2,christian,amp,jazz2,svs_unet,cunet,lasaft,google,meta}, to the best of our knowledge, there is no existing method that can directly address AMSS (see \S \ref{sec:related}).
This paper proposes a novel end-to-end neural network that performs AMSS according to the given textual query.
Designing a neural network for AMSS is straightforward if the sources of a given mixture track are observable. 
However, we assume that they are not observable because most audio data does not provide them in general.
In the assumed environment, modeling AMSS is very challenging because a sound object (e.g., a sample in a wave, a frequency bin in a spectrogram) is  `transparent'(\cite{trans}); a pixel in an image usually corresponds to only a single visual object, whereas a sound object carries information of multiple sources, as shown in Figure \ref{fig:challenge}.
Thus, we need different approaches for AMSS from the existing image manipulation techniques.

To address this challenge, we propose a neural network called AMSS-Net that extracts a feature map containing \textit{latent sources} (see \S \ref{sec:lcs_extract}) from the given mixture audio and selectively manipulates them while preserving irrelevant latent sources. 
We describe the AMSS-Net architecture in \S \ref{sec:amssnet}.

Another challenge is that existing datasets cannot be directly used for supervised training AMSS-Net.
If a training dataset of triples $\{(A^{(i)}, A'^{(i)}, S^{(i)})\}_{i=1}^{N}$, where $S^{(i)}$ is an AMSS description, $A^{(i)}$ is a mixture, and $A'^{(i)}$ is the manipulated audio according to $S^{(i)}$ is provided, we can train a neural network $net$ in a supervised manner by minimizing $\sum_{i=1}^{N}loss(net(A^{(i)}, S^{(i)}), A'^{(i)})$, where $loss$ is a distance metric such as $L_{2}$.
Unfortunately, there were no datasets currently available that directly target AMSS.
To address this issue, we propose a training framework for AMSS (see \S \ref{sec:training_framework}) that uses a `source observable multi-track dataset' such as Musdb18 \cite{musdb18}. 
To generate an AMSS triple on-the-fly, we apply audio transformations onto specific sources of a given multi-track using common methods from Digital Signal Processing (DSP) libraries.

We summarize our contributions as follows:

\begin{itemize}
    \item Our work is a pioneer study on selective audio manipulation.
    \item We propose a supervised training framework for AMSS based on source observable multi-track datasets and DSP libraries together with evaluation benchmarks for AMSS.
    \item We propose AMSS-Net, which performs multiple tasks and outperforms baselines on several tasks.

\end{itemize}

\section{Task Definition}
\label{sec:AMSS}

\subsection{Audio Manipulation on Specific Sources}

We define Audio Manipulation on Specific Sources (AMSS) as follows: for a given audio track $A$ and a given description $S$, AMSS aims to generate a manipulated audio track $A'$ that semantically matches $S$ while preserving contents in $A$ that are not described in $S$.
$A$ contains multiple sources, and $S$ describes the desired audio transformation and the targets, which we want to manipulate.
$S$ can be represented as a one-hot encoding or a textual query.
In this paper, we assume that $S$ is a textual query written in the Audio Manipulation Language described in \S \ref{sec:aml}.

As a proof-of-concept, this paper aims to verify that it is possible to train a neural network for AMSS. Specifically, we focus on modifying specified sources' sonic characteristics (e.g., loudness, panning, frequency content, and dereverberation). This paper does not address more complex manipulations such as distortion (see \S \ref{sec:discussion}).
Throughout the rest of the paper, we define an \textit{AMSS task} to be a set of instructions dealing with the same manipulation method.
Table \ref{table:amsstasks} lists nine AMSS tasks we try to model in this paper. We also define an \textit{AMSS task class} as a set of similar AMSS tasks.

\begin{table}[]
\centering

\begin{tabular}{|c|c|l|}
\hline
class                                                                                     & task  & DSP operations                   \\ \hline
\multirow{4}{*}{volume control}                                                             & separate     & masking the others \\ \cline{2-3}
       & mute         & masking targets \\  \cline{2-3}                                                                                               & increase vol & re-scaling (increase)               \\ \cline{2-3} 
                                                                                          & decrease vol & re-scaling (decrease)                \\ \hline
\multirow{2}{*}{\begin{tabular}[c]{@{}l@{}}volume control\\ (multi channel)\end{tabular}} & pan left     & re-scaling (left > mean > right) \\ \cline{2-3} 
                                                                                          & pan right    & re-scaling (left < mean < right)\\ \hline
\multirow{2}{*}{filter}                                                                   & lowpasss     & Low-pass Filter \\ \cline{2-3} 
                                                                                          & highpass     & High-pass Filter \\ \hline
delay                                                                                     & dereverb     & reverb$^{*}$         \\ \hline
\end{tabular}


    \caption{List of AMSS tasks modeled in this paper: ($*$) denotes reversed generation process (the line 5 in Algorithm \ref{alg:algorithm})}
    \label{table:amsstasks}

\end{table}

\subsection{Audio Manipulation Language}
\label{sec:aml}

We assume that $S$ is given as a textual query, such as ``apply light lowpass to drums''.
We make this assumption because textual querying enables us to naturally describe any pair of a transformation function and its target sources with detailed options.
For example, we can control the level of audio effects (which corresponds to the parameter settings of DSP functions) by simply inserting adjectives such as \textit{light}, \textit{medium}, or \textit{heavy} into the query.
It also can provide easy extensibility to natural language interfaces, which will be addressed in future works.

To this end, we propose an Audio Manipulation Language based on a probabilistic Context-Free Grammar (CFG) \cite{cfg} for AMSS.
Due to the page limit, we only present a subset of production rules (i.e., Rules (\ref{eq:desc})-(\ref{eq:srcs})) that define the query language's syntax for the \textit{filter class}.
The Full CFG is available online\footnote{https://kuielab.github.io/AMSS-Net/aml.html}.


\begin{subequations}
    \label{eq:cfg} 
    \begin{align}
        & <desc> \rightarrow <cls_{f}>   \label{eq:desc} \\
        & <cls_{f}> \rightarrow
            \, \textrm{\textit{\textbf{apply}}} \,\,
            <opt-filter> \,\,
            \textrm{\textit{\textbf{to}}} \,\,
            <srcs> \label{eq:cls}
        \\
        & <opt-filter> \rightarrow  \, <opt> \,\, <filter> \,\, | \, <filter> \label{eq:optfilter} \\
        & <opt> \rightarrow  \, \textrm{\textit{\textbf{light}}} \,\, | \textrm{\textit{\textbf{medium}}} \,\, | \textrm{\textit{\textbf{heavy}}}
        \label{eq:opt} \\
        & <filter> \rightarrow  \, \textrm{\textit{\textbf{lowpass}}} \,\, | \,\,  \textrm{\textit{\textbf{highpass}}}
        \label{eq:filter}
        \\
        & <srcs> \rightarrow  \, \textrm{\textit{\textbf{vocals}}} \,\, | \,\,  \textrm{\textit{\textbf{drums}}}
        \,\, | \,\,  \textrm{\textit{\textbf{bass}}}
        \,\, | \,\,  \textrm{\textit{\textbf{vocal, drums}}}
        \,\, | \,\, ... \label{eq:srcs}
    \end{align}
\end{subequations}

In the above rules, bold strings are terminal symbols, and strings enclosed in angle brackets are non-terminal symbols.
Each rule is of the form $A \rightarrow \alpha | \beta | ...$, which means that $A$ can be replaced with $\alpha$ or $\beta$.
In a CFG, we apply a rule to replace a single non-terminal symbol with one of the expressions.
Starting from the first symbol $<desc>$, we can generate a \textit{valid} query string by recursively applying rules until there is no non-terminal symbol.

For example, ``\textbf{\textit{apply medium lowpass to vocals, drums}}'' can derived from $<desc>$ by Rules (\ref{eq:desc})-(\ref{eq:srcs}).
We can also produce ``\textbf{\textit{apply lowpass to vocals, drums}}'' if we choose $<filter>$ instead of $<opt><filter>$. 
Since we set a default option for \textit{lowpass} level to be \textit{medium}, those two queries have the same meaning.

Rules (\ref{eq:cls})-(\ref{eq:filter}) are dependant on a AMSS task class, and Rule (\ref{eq:srcs}) is dependant on a given multi-track audio.
In this work, we use four AMSS task classes as shown in Table \ref{table:amsstasks}.
Since we use in the experiment  Musdb18\cite{musdb18} dataset of which track contains three named instruments (i.e., vocals, drums, bass), we set the right-hand side of Rule (\ref{eq:srcs}) to have all the possible permutations (13 expressions in total) 

\section{Related Work}
\label{sec:related}
Since the past decade, data-driven approaches for audio manipulation have been active research fields. 
Meanwhile, there were virtually no studies that directly addressed AMSS.
An exception was an early algorithmic method \cite{algorithmic}, which proposed a framework that identifies and controls volumes of desired sources based on digital signal processing.
Unlike \cite{algorithmic}, this paper presents a machine learning-based approach for various AMSS tasks as well as volume controls.
We review the literature related to machine learning-based studies for audio manipulation.

\textbf{Audio Effect Modeling}:
As described in \cite{christian}, audio effects are used to modify perceptual attributes of the given audio signal, such as loudness, spatialization, and timbre.
Recently, several methods \cite{marco0,marco1,marco2,christian,amp} have been proposed for audio effect modeling with deep neural networks.
\cite{marco0} proposed a convolutional network performing equalization (i.e., an audio effect that changes the harmonic and timbral characteristics of audio signals).
\cite{marco1,marco2} proposed convolutional and recurrent networks for nonlinear audio effects with Long and Short-Term Memory \cite{lstm}, such as distortion and Leslie speaker cabinet.
\cite{christian} presented an efficient neural network for modeling an analog dynamic range compressor enabling real-time operation on CPU.

All the methods above assume an audio input with a single source, while our method assumes a mixture of multiple sources as input.
Also, they are task-specific (the trained model provides only one audio effect), while our approach can perform several manipulations with a single model.

\textbf{Automatic Mixing}: 
\textit{Audio mixing} is a task that combines multi-track recordings into a single audio track.
Before taking the sum of signals, mix engineers usually apply various audio signal processing techniques such as equalization and panning for a better hearing experience.
Some approaches have attempted to automate audio mixing with deep neural networks.
For example, \cite{marco3} trained end-to-end neural networks for automatic mixing of drums, where the audio waveform of individual drum recordings is the input of the model and the post-produced stereo mix is the output.
\cite{christian1} proposed networks based on temporal dilated convolutions to learn neural proxies of specific audio effects and train them jointly to perform audio mixing.
These methods mainly focus on developing expert-type mixing models that combine individual sources into a mixture track, regardless of user control or input for the desired type of transformation. 
On the other hand, AMSS-Net aims to manipulate user-specified sources in a desired manner, preserving others, where individual sources are not observable.

Unlike \cite{marco3,marco4}, \cite{jazz1} assumed an environment where individual sources are not provided as input. They proposed an algorithmic framework that automatically remixes early jazz recordings, which are often perceived as irritating and disturbing from today’s perspective. It first decomposes the given input into individual tracks by means of acoustic source separation algorithms and remixes them using automatic mixing algorithms. \cite{jazz2} proposed a replacement of the source separation and the mixing processes by deep neural networks.  
The environment assumed in \cite{jazz2} is the most similar to ours.
However, \cite{jazz2} mainly focuses on remixing to change the audio mixing style, while this paper focuses on modifying specified sources in the desired manner.
Thus, \cite{jazz2} cannot directly address the AMSS problem.

\textbf{Source Separation}: 
Our work is also related to deep learning-based source separation methods \cite{svs_unet,tfctdf,phasen,dilatedlstm,cunet,lasaft,meta}. 
While early methods separate either a single source \cite{svs_unet,tfctdf,phasen} or multiple sources once \cite{dilatedlstm}, conditioned source separation methods \cite{cunet,lasaft,meta} isolate the source specified by an input symbol.
A conditioned source separation task can be viewed as an AMSS task where we want to simply \textit{mute} all the unwanted sources.
We propose a method that can perform several AMSS tasks as well as source separation tasks using a single model, and controlled with a textual query.

We can design an algorithmic framework for AMSS using an existing separation method as a preprocessor, similar to \cite{jazz1}.
Such a method can perform some AMSS tasks by separating the individual sources and applying the appropriate DSP functions to target sources. 
However, source separation methods also generate minor artifacts, which are not present in the original source.
Although they are negligible in each source, they can be large enough to be perceived when we sum the separated results for AMSS. 

Instead of such a hybrid framework, we propose an end-to-end neural network since our approach can model more complex transformations such as dereverberation, which cannot be easily modeled with DSP algorithms.
Still, our model must have the ability to extract the target sources to be manipulated. 
Inspired by recent deep learning-based methods, we design a novel model that extracts \textit{latent sources}, selectively manipulates them, aggregates them, and outputs a mixture that follows a textual query.

The concept of \textit{latent source} has been introduced in recent source separation methods \cite{google,lasaft}.
\cite{google} have trained their model to separate the given input into a variable number of latent sources, which can be remixed to approximate the original mixture. By carefully taking the weighted sum of separated latent sources, we can extract the desired source, such as clean speech.
\cite{lasaft} also use the concept of \textit{latent source} for conditioned source separation. 
They proposed the Latent Source-Attentive Frequency Transformation (LaSAFT) method, 
which extracts the feature map for each latent source and takes the weighted sum of them by using an attention mechanism.

We also use the concept of latent sources in AMSS-Net, where each latent source deals with a more detailed aspect of acoustic features than a symbolic-level source (e.g., `vocals').
Similar to \cite{lasaft}, we assume that a weighted sum of latent sources can represent a source, while \cite{google} assumed that latent sources are independent.
Unlike previous works, our approach is based on channel-level separation as described in \S \ref{sec:lcs_extract}.
Each decoding block of AMSS-Net extracts \textit{latent source channels} from the given feature map so that each channel can be correlated to a latent source. 

\begin{figure*}[h!]
  \includegraphics[width=\textwidth]{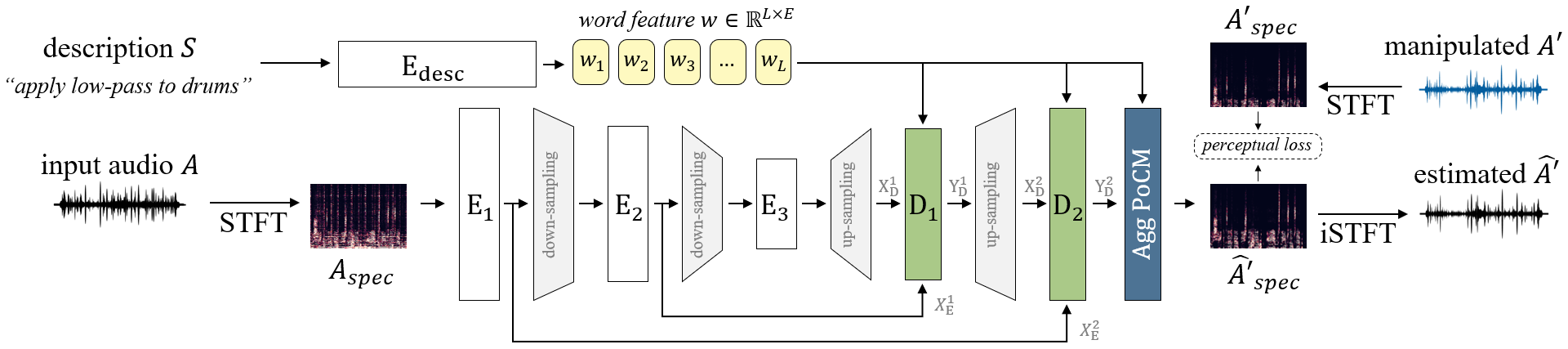}
  \caption{AMSS-Net Architecture}
  \label{fig:amssnet}
\end{figure*}

\section{Training Framework for AMSS}
\label{sec:training_framework}

We propose a novel training framework that uses a multi-track dataset. It generates an AMSS triple $(A, A', S)$ on the fly by applying DSP library transformations or audio effects onto target sources of a given multi-track audio file.
For instance, suppose that we have a randomly generated query string $S=$`apply lowpass to drums' using the CFG and a multi-track that consists of three sources, namely, a vocal track $a_{0}$, a bass track $a_{1}$, and a drum track $a_{2}$.
Our training framework takes the linear sum (i.e., $A=\sum a_{j}$) to generate $A$ to generate corresponding input $A$.
For the target audio $A'$, it computes $\sum_{j \notin \tau} a_{j} + \sum_{j \in \tau} f(a_{j}) $,  where $\tau$ and $f$ is the set of target sources and DSP function described in $S$, respectively. 
Our framework applies a Low-pass Filter (LPF) to $a_{2}$, and takes the sum for $A'$ as follows: $A'=a_{0} + a_{1} + LPF(a_{2})$. By doing so, it can generate an AMSS triple on-the-fly for a given description $S$.

For audio restoration tasks such as dereverberation, we swap A and A'.
For example, the framework applies reverb to $a_{2}$, takes the sum for $A'=a_{0} + a_{1} + reverb(a_{2})$, and returns $(A', A, S)$ instead of $(A, A', S)$ for the description  ``remove reverb from drums.''

\begin{algorithm}[htb]
\caption{AMSS Training Triple Generation}
\label{alg:algorithm}
\textbf{Input}: multi-track $\{a_{j}\}_{j=1}^{n}$,  a set $\mathcal{G}$ of triple generators \\
\textbf{Output}: a triple $(A, A', S)$
\begin{algorithmic}[1]
\STATE randomly sample a generator $g$ from $\mathcal{G}$
\STATE $S, f, \tau  \leftarrow g.gen\_random\_generate()$
\STATE $A \leftarrow \sum _{j=1}^{n} a_{j}$
\STATE $A' \leftarrow \sum_{j \notin \tau} a_{j} + \sum_{j \in \tau} f(a_{j}) $

\IF {g is for removing effect}
\STATE $(A, A', S) \leftarrow (A', A, S)$
\ENDIF
\STATE \textbf{return} $(A', A, S)$

\end{algorithmic}
\end{algorithm}

The training framework has a set $\mathcal{G}$ of triple generators. A generator $g\in \mathcal{G}$ has a subset of CFG for text query generation and the corresponding DSP function $f$, which is used for computing $A'$. $g$ also has an indicator that describes whether $g$ is for applying or removing the effect. For a multi-track $\{A_{j}\}_{j=1}^{n}$ and a set $\mathcal{G}$ of triple generators, our framework generates an AMSS triple by using Algorithm \ref{alg:algorithm}, where $n$ denotes the number of sources.

\section{AMSS-Net Architecture}
\label{sec:amssnet}

Our AMSS-Net takes as input an audio track $A$ with a text description $S$ and generates the manipulated audio track $\hat{A'}$ as shown in Figure \ref{fig:amssnet}.
It consists of two sub-networks, i.e., a Description Encoder $\textrm{E}_{desc}$ and a Spectrogram Encoder-Decoder Network $SEDN$.
$\textrm{E}_{desc}$ extracts word features $w \in \mathbb{R}^{L \times E}$ from $S$, where $E$ denotes the dimension of the word features and $L$ denotes the number of words, to analyze the meaning of $S$.
$SEDN$ takes as input word features $w$ and the complex-valued spectrogram $A_{spec}$ of the input audio $A$. 
Conditioned on the word feature $w$, it estimates the complex-valued spectrogram $\hat{A'}_{spec}$, from which $\hat{A'}$ can be reconstructed using iSTFT.
We train the AMSS-Net by minimizing the $L_2$ loss between the ground-truth spectrogram $A'_{spec}$ of $A'$ and the estimated $\hat{A'}_{spec}$.

\subsection{Description Encoder}
\label{sec:de}

The description encoder $\textrm{E}_{desc}$ encodes the given text description $S$ written in the Audio Manipulation Language (\S \ref{sec:aml}) to word features $w \in \mathbb{R}^{L \times E}$, where we denote the dimension of each word feature by $E$ and the number of words in $S$ by $L$.
It embeds each word to a dense vector using a word embedding layer and then encodes the embedded description using a bidirectional Recurrent Neural Network (Bi-RNN) \cite{birnn}.
We use pre-trained word embeddings such as GloVe\cite{glove} to initialize the weight of the embedding layer since they were trained to capture the syntactic and semantic meaning of words.

\subsection{Spectrogram Encoder-Decoder Network}
\label{sec:cunet}

The Spectrogram Encoder-Decoder Network $SEDN$ estimates the complex-valued spectrogram $\hat{A'}_{spec}$, conditioned on the extracted word features $w \in \mathbb{R}^{L \times E}$.
It is an encoder-decoder network that has the same number of down-sampling layers and up-sampling layers as depicted in Figure \ref{fig:amssnet}.
It extracts down-sampled representations from $A_{spec}$ in the encoding phase and generates up-sampled representations in the decoding phase.
The output of the last decoding block is fed to the \textit{Aggregate PoCM} (see \S \ref{sec:aggregate_pocm}) that generates the output $\hat{A'}_{spec}$.

As illustrated in Figure \ref{fig:amssnet}, it has direct connections between the encoding blocks and their counterpart decoding blocks, which help decoding blocks recover fine-grained details of the target.
Instead of concatenation or summation commonly used in several U-Net-based architectures \cite{svs_unet,tfctdf,cunet,lasaft}, we propose a Channel-wise Skip Attention (CSA) mechanism (\S \ref{sec:csa}) that attentively aggregates latent source channels to reconstruct the original channels.

As shown in Figure \ref{fig:amssnet}, $SEDN$ consists of several components: \textit{encoding blocks}, \textit{down/up-sampling layers}, \textit{decoding blocks}, and an \textit{Aggregate PoCM}.
We use strided convolutions and transposed convolutions for down-sampling and up-sampling, respectively.
We introduce other componets in \S \ref{sec:encoding}, \S \ref{sec:decoding}, and \S \ref{sec:aggregate_pocm}.

\subsection{Spectrogram Encoding Block}
\label{sec:encoding}

$SEDN$ uses multiple encoding blocks in the encoding phase to capture common sonic properties residing in the input spectrogram.
The $k^{th}$ encoding block $E_{k}$ transforms an input spectrogram-like tensors into an equally-sized tensor.
We adopt the encoding block called TFC-TDF proposed in \cite{tfctdf}, which applies densely-connected 2-d convolutions to the given spectrogram-like representations followed by a fully connected layer that enhances features of frequency patterns observed in the frequency axis.

\begin{figure*}[h!]
   \includegraphics[width=\textwidth]{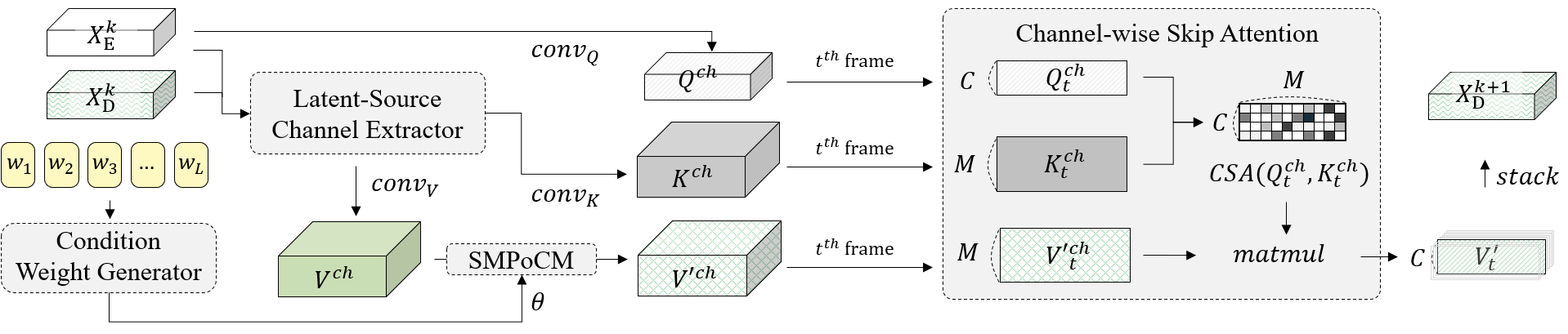}
  \caption{$k^{th}$ decoding block ($D_k$)}
  \label{fig:decoder}
\end{figure*}

\subsection{Spectrogram Decoding Block}
\label{sec:decoding}

In the decoding phase, $SEDN$ uses multiple decoding blocks.
Each decoding block first extracts a feature map in which each channel corresponds to a specific latent source. It selectively manipulates them conditioned on the AMSS description and aggregates channels using a channel-wise attention mechanism to minimize information loss during channel reconstruction.

As shown in Figure \ref{fig:decoder}, the $k^{th}$ decoding block $D_k$ takes three inputs: (1) $X_{D}^{k}$: features from the previous decoding block, (2) $X_{E}^{k}$: features from the skip connection and (3) $w\in\mathbb{R}^{L\times E}$: word features. The first block takes the up-sampled features from the encoder instead because it has no previous decoding block.

Each decoding block consists of four components: Latent Source Channel (LSC) Extractor, Condition weight generator, Selective Manipulation via PoCMs (SMPoCM), and Channel-wise Skip Attention (CSA).
We illustrate the overall workflow in Figure \ref{fig:decoder}.



\subsubsection{Latent Source Channel Extractor}
\label{sec:lcs_extract}
\begin{figure}[t]
    \centering
    \includegraphics[width=\columnwidth]
    {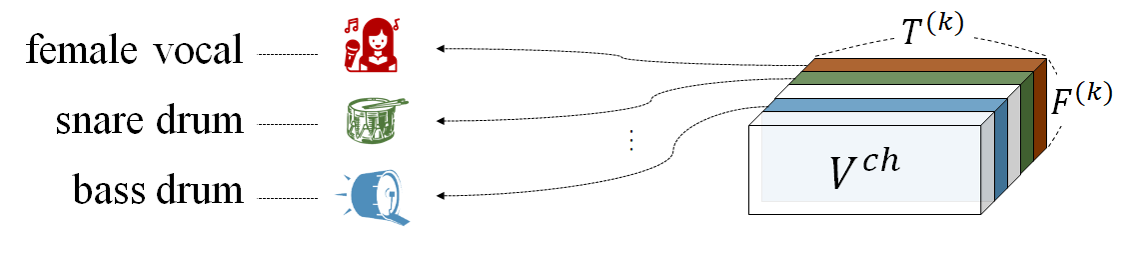}
    \caption{Conceptual View of Latent Source Channels}
    \label{fig:latentsource}
\end{figure}

We assume that we can learn representations of latent sources that deal with more detailed acoustic features than symbolic-level sources.
The Latent Source Channel (LSC) extractor aims to generate a feature map $V^{ch}$, in which each channel deals with a latent source.
Figure \ref{fig:latentsource} visualizes the conceptual view of latent source channels.
Each channel is a spectrogram-like representation of size $T^{(k)} \times F^{(k)}$ dealing with a specific latent source.
For example, the blue channel in Figure \ref{fig:latentsource} deals with the acoustic features observed in the bass drum.
We can also generate an audio track from a single latent source channel.
In \S \ref{sec:latentsource}, we visualize and discuss results of the audio generation.

The LSC extractor takes $X_{E}^{k}, X_{D}^{k} \in \mathbb{R}^{C \times T^{(k)} \times F^{(k)}}$ as input and extracts feature maps with $M$ latent source channels, where $C$ refers to the number of channels, $T^{(k)} \times F^{(k)}$ refers to the shape of the spectrogram-like features, and $M$ denotes the number of latent sources. 
It first concatenates $X_{E}^{k}$ and $X_{D}^{k}$ to obtain $[X_{E}^{k};X_{D}^{k}]\in \mathbb{R}^{2C \times T^{(k)} \times F^{(k)}}$, and applies a TFC-TDF block \cite{tfctdf} to $[X_{E}^{k};X_{D}^{k}]$ to obtain acoustic features $X^k$ for latent source separation.

To extract a feature map $V^{ch}\in \mathbb{R}^{M \times T^{(k)} \times F^{(k)}}$ with $M$ latent source channels, it applies a $1\times1$ convolution to $X^{k}$.
We denote this convolution as $conv_V$ since its role is a \textit{value generator} in the context of the channel-wise skip attention mechanism \S \ref{sec:csa} as shown in Figure \ref{fig:decoder}.
Similar to $conv_V$, we apply another $1\times1$ convolution called $conv_K$ to $X^k$, of which role is a \textit{key generator}, to obtain $K^{ch}$.
By the guide of the channel-wise skip attention mechanism, the LSC extractor is expected to extract \textit{latent source channels} from the mixture features so that each channel deals with a latent source.

\subsubsection{Selective Manipulation via PoCMs}

Since we have to manipulate specific features while preserving other features, selective manipulation requires a more sophisticated modulation than existing methods such as Feature-wise Linear Modulation (FiLM) \cite{film}, or Point-wise Convolutional Modulation (PoCM) \cite{lasaft}.

We propose SMPoCM, an extension of PoCM, that aims to manipulate features for the given AMSS task selectively.
As shown in Figure \ref{fig:decoder}, it takes  as input $V^{ch}$ and condition parameters $\theta=(\theta_s, \theta_m, \theta_i)$, generated by the LSC extractor and the condition weight generator, respectively. It outputs ${V'}^{ch}\in \mathbb{R}^{M \times T^{(k)} \times F^{(k)}}$, a selectively manipulated feature. 

Inspired by Long Short-term Memory (LSTM) \cite{lstm}, the SMPoCM uses three different PoCMs: (1) a selective gate PoCM with $\theta_{s}$ to determine how much we should manipulate each latent source channel, (2) a manipulation PoCM with $\theta_{m}$ to manipulate specific features, and (3) an input gate PoCM with $\theta_{i}$ to determine how much we should emit the manipulated features.

Before defining SMPoCM formally, we summarize the behavior of PoCM as follows: a PoCM is a point-wise convolution (i.e., $1\times1$ convolution) of which the condition weight generator provides parameters.
We now give a definition of SMPoCM as follows: $\mathrm{SMPoCM}(X|\theta)$ = $i$ $\odot$ $tanh$ $(\mathrm{PoCM}(s \odot X, \theta_{m}))$ $+(1-s)$ $\odot$ $X$
, where $s$ is defined as $ \sigma(\mathrm{PoCM}(X, \theta_{s}))$, $i$ is defined as $ \sigma(\mathrm{PoCM}(X, \theta_{i}))$, $\odot$ is Hadamard product, and $ \sigma$ is a sigmoid function.
The total number of parameters in $\theta$ is about $3(M^{2}+M)$ (i.e., three point-wise convolutions).

SMPoCM naturally models the selective modulation required for modeling AMSS tasks. For example, if $i^{th}$ latent source should be preserved for the given input, then the $i^{th}$ channel of $s$ would be trained to have near-zero values.

\subsubsection{Condition Weight Generator}
\label{sec:wg}

Given $w \in \mathbb{R}^{L \times E}$, the condition weight generator generates parameters $\theta=(\theta_{s}, \theta_{m}, \theta_{i})$.
We exploit the attention mechanism to determine which word should be attended to $\theta_{s}$, $\theta_{m}$, and $\theta_{i}$ respectively.

The weight generator has a learnable matrix $\Theta \in \mathbb{R}^{3\times d_{k}}$, where we denote the cardinality of each PoCM task embedding by $d_{k}$.
It also has two linear layers $linear_{k}$ and $linear_{v}$ that embed $w$ to $w_{key} \in \mathbb{R}^{L \times d_{k}}$ and $w_{value} \in \mathbb{R}^{L \times d_{k}}$ respectively. To determine which word we should attend for each task, we compute the scaled attention \cite{transformer} as follows: $\alpha^{wg} = $ $softmax(\frac{\Theta w_{key}}{ \sqrt{d_{k}} })w_{value}^T$. Finally, it generates 
$\theta_{s}$, $\theta_{m}$, and $\theta_{i}$ as follows: $\theta_{s}=linear_{s}(\alpha^{wg}[0,:])$, 
$\theta_{s}=linear_{m}(\alpha^{wg}[1,:])$, and
$\theta_{s}=linear_{i}(\alpha^{wg}[2,:])$, where $linear_{s}$, $linear_{m}$, and $linear_{i}$ are fully-connected layers.

\subsubsection{Channel-wise Skip Attention}
\label{sec:csa}

Inspired by skip attention \cite{sa} and \cite{ca}, we propose a Channel-wise Skip Attention (CSA) mechanism.
It attentively aggregates $M$ latent source channels to restore the number of channels to the same as the input (i.e., $C$).
The goal of CSA is to minimize information loss during channel reconstruction to preserve other features that are irrelevant to the description.

Figure \ref{fig:decoder} overviews the workflow required to prepare the query feature $Q^{ch}$, key feature $K^{ch}$, and the value feature $V'^{ch}$.
To obtain $Q^{ch} \in \mathbb{R}^{C \times T^{(k)} \times F^{(k)}}$, we apply $conv_Q$, a $1\times1$ convolution to $X_{E}^{k}$. 

For each frame, CSA aims to capture channel-to-channel dependencies between $X_{E}^{k}$ that encodes the original acoustic features of $A$ and the feature map $K^{ch}$ of isolated latent sources obtained by the LSC extractor.
It is worth noting that we use $K^{ch}$ for computing attention matrix instead of $V'^{ch}$ since $V'^{ch}$, which SMPoCM modulated, no longer has the same information as $X_{E}^{k}$. 

For $Q^{ch}_{t}=Q^{ch}[:,t,:]$ and $K^{ch}_{t}=K^{ch}[:,t,:]$, we compute the scaled dot product attention matrix \cite{transformer} as follows:

\begin{equation}
    CSA(Q^{ch}_{t}, K^{ch}_{t}) = softmax(\frac{Q^{ch}_{t}(K^{ch}_{t})^{T}}{\sqrt{F^{(k)}}})
\end{equation}

The attention weight $CSA(Q^{ch}_{t},K^{ch}_{t})_{i,j}$ represents the correlation between the $i^{th}$ channel of the original audio features and the $j^{th}$ latent source channel of the decoded audio features. 
Finally, we can obtain the decoding block's output $y_{D}^{k}$, where $y_{D}^{k}[:,t,:]$ is defined as $CSA(Q^{ch}_{t}, K^{ch}_{t})V'^{ch}[:,t,:]$.

\subsection{Aggregate PoCM}
\label{sec:aggregate_pocm}
The Aggregate Pocm is similar to the SMPoCM other than two key differences. First, the Aggregate PoCM only has one PoCM that is not followed by any activation. Second, the Aggregate PoCM reduces the number of channels from $C$ to $4$ (i.e., the number of channels of the two-channeled complex-valued spectrogram) while the SMPoCM's input and output have the same number of channels. 

\section{Experiments}

\label{sec:exp}

\subsection{Experiment Setup}

We evaluate the proposed model both qualitatively and quantitatively on various AMSS tasks described in Table \ref{table:amsstasks} using the Musdb18 dataset. We compare its performance with baselines to verify our architecture.

\subsubsection{Training Framework}

Musdb18 \cite{musdb18} contains 86 tracks for training, 14 tracks for validation, and 50 tracks for the test. Each track is stereo, sampled at 44100 Hz, and each data tuple consists of the mixture and four sources: vocals, drums, bass, and other. 
We implemented the training framework based on Musdb18 and pysndfx\footnote{https://pypi.org/project/pysndfx/}, a Python DSP library. We train all models based on this framework with 9 AMSS tasks listed in Table \ref{table:amsstasks}. 
For each task, we implement triple generators based on the Audio Manipulation Language (\S \ref{sec:aml}) and Musdb18. We exclude `other' since it is not a single instrument. With the set $\mathcal{G}$ of triple generators and randomly generated multi-tracks obtained by data augmentation \cite{blend}, we generate AMSS triples using Algorithm \ref{alg:algorithm} for training.

\subsubsection{Training Environment}

We train models using Adam \cite{adam} with learning rate  $lr \in [0.0001, 0.001]$. Starting with a learning rate $lr$, we halved $lr$ and restarted training when the current $lr$ seemed to be too high.  
Each model is trained to minimize the $L_2$ loss between the ground-truth and estimated spectrograms.
For validation, we use the $L_1$ loss of target and estimated signals. It takes about two weeks to converge when we train models with a single 2080Ti GPU.

\subsection{Model Configurations}

To validate the effectiveness of AMSS-Net, we compared it with the two baselines. One does not use CSA (AMSS-Net w/o CSA), and the other does not use SMPoCM in decoding blocks (AMSS-Net w/o SM). The baseline model without CSA uses an LSC extractor with LaSAFT block \cite{lasaft} instead of TFC-TDF \cite{tfctdf} to compensate the absence of CSA.
The model without SMPoCM uses a single PoCM with tanh activation in its decoding block.
An AMSS-Net has about 4.3M, a baseline without CSA has about 4.9M, and a baseline without SMPoCM has about 2.4M.

For hyper-parameter setting, we use a similar configuration of models with an FFT window size of 2048 in \cite{lasaft}.
Every model has three encoding blocks, two decoding blocks, an additional Aggregate PoCM block.
We assume that there are eight latent sources (i.e., $M=8$). We also adopt the multi-head attention mechanism \cite{transformer} for CSA, where we set the number of heads to 6.
The STFT parameter of each model is as follows: an FFT window size of 2048 and a hop size of 1024.


\begin{table}[]
\begin{tabular}{|l|c|c|c|c|c|}
\hline
model             & vocals & drums & bass & other & AVG \\ \hline
Meta-TasNet & 6.40   & 5.91  & 5.58 & 4.19 & 5.52    \\ \hline

LaSAFT-GPoCM-Net$_{12}$ & \textbf{7.33}  & 5.68  & \textbf{5.63} & \textbf{4.87}  & \textbf{5.88}    \\ \hline

LaSAFT-GPoCM-Net$_{11}$ & 6.96  & 5.84  & 5.24 & 4.54  & 5.64    \\ \hline 

AMSS-Net$_{separate}$ 
    & \begin{tabular}[c]{@{}c@{}} 6.78\\ $\pm$ .12 \end{tabular}
    & \begin{tabular}[c]{@{}c@{}} \textbf{5.92}\\ $\pm$ .03 \end{tabular}
    & \begin{tabular}[c]{@{}c@{}} 5.10\\ $\pm$ .06 \end{tabular}
    & \begin{tabular}[c]{@{}c@{}} 4.51\\ $\pm$ .10 \end{tabular}
    & \begin{tabular}[c]{@{}c@{}} 5.58\\ $\pm$ .90 \end{tabular}
    \\ \hline  \hline
    
AMSS-Net
    & \begin{tabular}[c]{@{}c@{}} \textbf{6.34}\\ $\pm$ .016 \end{tabular}
    & \begin{tabular}[c]{@{}c@{}} 5.53\\ $\pm$ .07 \end{tabular}
    & \begin{tabular}[c]{@{}c@{}} \textbf{4.33}\\ $\pm$ .13 \end{tabular}
    & \begin{tabular}[c]{@{}c@{}} \textbf{3.99}\\ $\pm$ .07 \end{tabular}
    & \begin{tabular}[c]{@{}c@{}} \textbf{5.05}\\ $\pm$ .99 \end{tabular}
    \\ \hline
\quad w/o CSA    
    & \begin{tabular}[c]{@{}c@{}} 6.03\\ $\pm$ .24 \end{tabular}
    & \begin{tabular}[c]{@{}c@{}} 5.53\\ $\pm$ .12 \end{tabular}
    & \begin{tabular}[c]{@{}c@{}} 4.22\\ $\pm$ .09 \end{tabular}
    & \begin{tabular}[c]{@{}c@{}} 3.73\\ $\pm$ .23 \end{tabular}
    & \begin{tabular}[c]{@{}c@{}} 4.88\\ $\pm$ .99 \end{tabular}
    \\ \hline

\quad w/o SMPoCM
    & \begin{tabular}[c]{@{}c@{}} 6.08\\ $\pm$ .25 \end{tabular}
    & \begin{tabular}[c]{@{}c@{}} \textbf{5.63}\\ $\pm$ .08 \end{tabular}
    & \begin{tabular}[c]{@{}c@{}} 4.28\\ $\pm$ .19 \end{tabular}
    & \begin{tabular}[c]{@{}c@{}} 3.76\\ $\pm$ .04 \end{tabular}
    & \begin{tabular}[c]{@{}c@{}} 4.94\\ $\pm$ .99 \end{tabular}
    \\ \hline

    
\end{tabular}
    \caption{A comparison SDR performance. LaSAFT-GPoCM-Net$_x$ uses FFT window size of $2^{x}$ }
    \label{table:task1}
\end{table}

\begin{table*}[!t]
\centering
\begin{tabular}{|l|c|c|c|c|c|c|c|c|c|c|c|c|}
\hline
\multicolumn{1}{|c|}{{\textit{}}}
    & \multicolumn{3}{c|}{pan left}                          
    & \multicolumn{3}{c|}{pan right} 
    & \multicolumn{3}{c|}{decrease volume}
    & \multicolumn{3}{c|}{increase volume} \\ \cline{2-13}
    
    & voc & drum & bass 
    & voc & drum & bass 
    & voc & drum & bass 
    & voc & drum & bass 
\\ \hline 

reference loss        & 4.62 & 8.20 & 2.18 &
4.66 & 8.30 & 2.13 &
4.06 & 6.99 & 2.00 &
6.68 & 9.70 & 3.48 
\\ \hline




AMSS-Net 
    
& \begin{tabular}[c]{@{}c@{}} \textbf{3.32} \\ $\pm$0.19 \end{tabular} & \begin{tabular}[c]{@{}c@{}} \textbf{4.81} \\ $\pm$0.16 \end{tabular} & \begin{tabular}[c]{@{}c@{}} \textbf{2.11} \\ $\pm$0.13 \end{tabular} & \begin{tabular}[c]{@{}c@{}} \textbf{3.34} \\ $\pm$0.19 \end{tabular} & \begin{tabular}[c]{@{}c@{}} \textbf{4.85} \\ $\pm$0.19 \end{tabular} & \begin{tabular}[c]{@{}c@{}} \textbf{2.11} \\ $\pm$0.17 \end{tabular} & \begin{tabular}[c]{@{}c@{}} \textbf{2.72} \\ $\pm$0.23 \end{tabular} & \begin{tabular}[c]{@{}c@{}} \textbf{3.78} \\ $\pm$0.18 \end{tabular} & \begin{tabular}[c]{@{}c@{}} \textbf{1.93} \\ $\pm$0.13 \end{tabular} & \begin{tabular}[c]{@{}c@{}} \textbf{3.49} \\ $\pm$0.2 \end{tabular} & \begin{tabular}[c]{@{}c@{}} \textbf{4.36} \\ $\pm$0.18 \end{tabular} & \begin{tabular}[c]{@{}c@{}} \textbf{2.9} \\ $\pm$0.19 \end{tabular}

    \\ \hline
\quad w/o CSA    
& \begin{tabular}[c]{@{}c@{}} 3.65 \\ $\pm$0.17 \end{tabular} & \begin{tabular}[c]{@{}c@{}} 5.46 \\ $\pm$0.2 \end{tabular} & \begin{tabular}[c]{@{}c@{}} 2.61 \\ $\pm$0.33 \end{tabular} & \begin{tabular}[c]{@{}c@{}} 3.63 \\ $\pm$0.11 \end{tabular} & \begin{tabular}[c]{@{}c@{}} 5.53 \\ $\pm$0.23 \end{tabular} & \begin{tabular}[c]{@{}c@{}} 2.58 \\ $\pm$0.3 \end{tabular} & \begin{tabular}[c]{@{}c@{}} 3 \\ $\pm$0.13 \end{tabular} & \begin{tabular}[c]{@{}c@{}} 4.26 \\ $\pm$0.14 \end{tabular} & \begin{tabular}[c]{@{}c@{}} 2.32 \\ $\pm$0.33 \end{tabular} & \begin{tabular}[c]{@{}c@{}} 4.47 \\ $\pm$0.12 \end{tabular} & \begin{tabular}[c]{@{}c@{}} 5.11 \\ $\pm$0.18 \end{tabular} & \begin{tabular}[c]{@{}c@{}} 3.66 \\ $\pm$0.4 \end{tabular}

    \\ \hline

\quad w/o SMPoCM  
& \begin{tabular}[c]{@{}c@{}} 4.34 \\ $\pm$0.2 \end{tabular} & \begin{tabular}[c]{@{}c@{}} 5.51 \\ $\pm$0.16 \end{tabular} & \begin{tabular}[c]{@{}c@{}} 3.23 \\ $\pm$0.19 \end{tabular} & \begin{tabular}[c]{@{}c@{}} 4.23 \\ $\pm$0.18 \end{tabular} & \begin{tabular}[c]{@{}c@{}} 5.5 \\ $\pm$0.11 \end{tabular} & \begin{tabular}[c]{@{}c@{}} 3.19 \\ $\pm$0.17 \end{tabular} & \begin{tabular}[c]{@{}c@{}} 3.55 \\ $\pm$0.16 \end{tabular} & \begin{tabular}[c]{@{}c@{}} 4.41 \\ $\pm$0.16 \end{tabular} & \begin{tabular}[c]{@{}c@{}} 2.92 \\ $\pm$0.17 \end{tabular} & \begin{tabular}[c]{@{}c@{}} 4.16 \\ $\pm$0.14 \end{tabular} & \begin{tabular}[c]{@{}c@{}} 5.2 \\ $\pm$0.15 \end{tabular} & \begin{tabular}[c]{@{}c@{}} 3.4 \\ $\pm$0.17 \end{tabular}

    \\ 

\end{tabular}

\begin{tabular}{|l|c|c|c|c|c|c|c|c|c|c|c|c|}
\hline
\multicolumn{1}{|c|}{{\textit{}}}
    & \multicolumn{3}{c|}{lowpass filter}
    & \multicolumn{3}{c|}{highpass filter}
    & \multicolumn{3}{c|}{dereverberation} 
    & \multicolumn{3}{c|}{mean} \\ \cline{2-13}
& voc & drum & bass 
& voc & drum & bass 
& voc & drum & bass 
& voc & drum & bass 
\\ \hline

reference loss    
& 9.02
& 16.27
& 1.12
& 6.72
& 8.22
& 4.74
& 9.43
& 11.07
& 5.61
& 6.46
& 9.82
& 3.04
\\ \hline




AMSS-Net 

& \begin{tabular}[c]{@{}c@{}} \textbf{7.32} \\ $\pm$0.19 \end{tabular} & \begin{tabular}[c]{@{}c@{}} \textbf{10.92} \\ $\pm$0.15 \end{tabular} & \begin{tabular}[c]{@{}c@{}} \textbf{2.65} \\ $\pm$0.09 \end{tabular} & \begin{tabular}[c]{@{}c@{}} 5.65 \\ $\pm$0.28 \end{tabular} & \begin{tabular}[c]{@{}c@{}} \textbf{6.9} \\ $\pm$0.3 \end{tabular} & \begin{tabular}[c]{@{}c@{}} \textbf{3.8} \\ $\pm$0.11 \end{tabular} & \begin{tabular}[c]{@{}c@{}} \textbf{6.12} \\ $\pm$0.29 \end{tabular} & \begin{tabular}[c]{@{}c@{}} \textbf{6.72} \\ $\pm$0.23 \end{tabular} & \begin{tabular}[c]{@{}c@{}} \textbf{4.18} \\ $\pm$0.24 \end{tabular} & \begin{tabular}[c]{@{}c@{}} \textbf{4.57} \\ $\pm$1.70 \end{tabular} & \begin{tabular}[c]{@{}c@{}} \textbf{6.05} \\ $\pm$2.33 \end{tabular} & \begin{tabular}[c]{@{}c@{}} \textbf{2.81} \\ $\pm$0.85 \end{tabular}

    \\ \hline
\quad w/o CSA    
& \begin{tabular}[c]{@{}c@{}} 7.44 \\ $\pm$0.07 \end{tabular} & \begin{tabular}[c]{@{}c@{}} 12.37 \\ $\pm$0.5 \end{tabular} & \begin{tabular}[c]{@{}c@{}} 3.15 \\ $\pm$0.21 \end{tabular} & \begin{tabular}[c]{@{}c@{}} \textbf{5.49 }\\ $\pm$0.25 \end{tabular} & \begin{tabular}[c]{@{}c@{}} 7.41 \\ $\pm$0.67 \end{tabular} & \begin{tabular}[c]{@{}c@{}} 4.23 \\ $\pm$0.4 \end{tabular} & \begin{tabular}[c]{@{}c@{}} 6.18 \\ $\pm$0.12 \end{tabular} & \begin{tabular}[c]{@{}c@{}} 7.08 \\ $\pm$0.07 \end{tabular} & \begin{tabular}[c]{@{}c@{}} 4.28 \\ $\pm$0.1 \end{tabular} & \begin{tabular}[c]{@{}c@{}} 4.84 \\ $\pm$1.53 \end{tabular} & \begin{tabular}[c]{@{}c@{}} 6.75 \\ $\pm$2.60 \end{tabular} & \begin{tabular}[c]{@{}c@{}} 3.26 \\ $\pm$0.83 \end{tabular}

    \\ \hline

\quad w/o SMPoCM 
& \begin{tabular}[c]{@{}c@{}} 8.11 \\ $\pm$0.1 \end{tabular} & \begin{tabular}[c]{@{}c@{}} 12.06 \\ $\pm$0.31 \end{tabular} & \begin{tabular}[c]{@{}c@{}} 3.67 \\ $\pm$0.23 \end{tabular} & \begin{tabular}[c]{@{}c@{}} 6.81 \\ $\pm$0.21 \end{tabular} & \begin{tabular}[c]{@{}c@{}} 7.84 \\ $\pm$0.24 \end{tabular} & \begin{tabular}[c]{@{}c@{}} 4.85 \\ $\pm$0.42 \end{tabular} & \begin{tabular}[c]{@{}c@{}} 6.68 \\ $\pm$0.12 \end{tabular} & \begin{tabular}[c]{@{}c@{}} 7.17 \\ $\pm$0.24 \end{tabular} & \begin{tabular}[c]{@{}c@{}} 4.68 \\ $\pm$0.05 \end{tabular} & \begin{tabular}[c]{@{}c@{}} 5.41 \\ $\pm$1.67 \end{tabular} & \begin{tabular}[c]{@{}c@{}} 6.81 \\ $\pm$2.47 \end{tabular} & \begin{tabular}[c]{@{}c@{}} 3.71 \\ $\pm$0.75 \end{tabular} 

    \\ \hline
\end{tabular}

\caption{A RMSE-MFCC Comparison of our models with baselines, over 7 AMSS tasks applied to vocals, drums, and bass}
\label{table:task2}

\end{table*}

\subsection{Quantitative Analysis}

\subsubsection{Evaluation of separate and mute tasks (Table \ref{table:task1})}
To evaluate \textit{separate} and \textit{mute} tasks, we use Source-to-Distortion (SDR) \cite{bss} metric by using the official tool\footnote{https://github.com/sigsep/sigsep-mus-eval}. 
The Musdb18 tasks are separating vocals, drums, bass, and other, which corresponds to the following AMSS: `separate vocals,' `separate drums,' `separate bass,' and 'mute vocals, drums, bass.' 
Using the SDR metric for these two tasks allows us to compare our model's source separation performance with other state-of-the-art models for conditioned source separation.
Following the official guideline, we report the median SDR value over all the test set tracks for each run and report the mean SDR over three runs (with a different random seed).

Table \ref{table:task1} summarizes the result, where AMSS-Net shows comparable or and slightly inferior performance compared to state-of-the-art conditioned separation models, namely, Meta-TasNet \cite{meta} and LaSAFT-GPoCM-Nets. AMSS-Net outperforms the baselines for all sources except for drums, where the gap is not significant. It is worth noting that ours can perform other AMSS tasks and show promising results on source separation tasks. If we train AMSS-Net solely for Musdb18 tasks (AMSS-Net$_{separate}$), then it shows comparable performance to LaSAFT-GPoCM-Net$_{11}$ whose FFT window size is the same as ours.

\subsubsection{Evaluation of other AMSS tasks (Table 3)}

Since there is no reference evaluation metric for AMSS, we propose the new evaluation benchmark.
The benchmark script is available online\footnote{https://github.com/kuielab/AMSS-Net/blob/main/task2\_eval.py}.
For evaluation metric, we extract Mel-frequency Cepstral coefficients (MFCC) for $A'$ and $\hat{A}'$, and compute the Root Mean Square Error (RMSE) of them, since MFCC approximates the human perception of the given track. We refer to this metric as RMSE-MFCC.
We report the mean RMSE-MFCC value over all the test set tracks for each run and report the mean RMSE-MFCC over three runs.

Table \ref{table:task2} summarizes the results, where \textit{reference loss} is the RMSE-MFCC of $A$ and $A'$. Reference loss provides information about the amount of manipulation needed to model each AMSS task. As described in Table \ref{table:task2}, AMSS-Net outperforms all the AMSS task but a task of ``highpass filter to vocals''. 
Significantly, the model without SMPoCM is inferior to AMSS-Net for every task. 
It indicates that SMPoCM significantly contributes to improving the quality of AMSS results.
CSA also improves the performance of AMSS-Net for most of the AMSS tasks. 
CSA might degrade the performance for more difficult AMSS tasks by forcing the model to over-correlate the latent source channels with the mixture channels.
However, it reduces artifacts created during progressive manipulation as described in \S \ref{sec:progresive}

\subsection{Latent Source Channels}
\label{sec:latentsource}

As mentioned in the LSC extractor (\S \ref{sec:lcs_extract}), we design AMSS-Net to perform latent source-level analysis.
Such analysis enables AMSS-Net to perform delicate manipulation for the given AMSS task.
To verify that AMSS-Net decoding blocks can extract a feature map in which each channel corresponds to a specific latent source, we generate an audio track from a single latent source channel.
We mask all channels in the manipulated feature map $V'^{ch}$ except for a single latent source channel and fed it to the remaining sub-networks to generate the audio track during the last decoding block.

\begin{figure}[h]

\centering
\begin{tabular}{cc}
    \includegraphics[width=40mm]{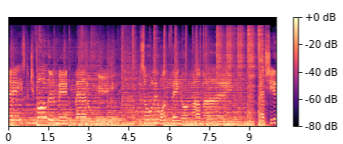} &
    \includegraphics[width=40mm]{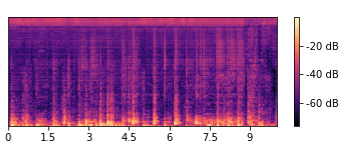}    \\
    (a) original & (b) kick drum \\
    \includegraphics[width=40mm]{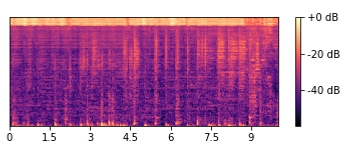} &
    \includegraphics[width=40mm]{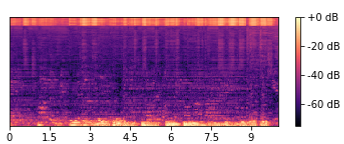}    \\
    (c) high hat  & (d) non-percussion \\
\end{tabular} 

\caption{Mel-Spectrogram of single latent-source channel}
\label{fig:mel_spec_latent_source}
\end{figure}

Figure \ref{fig:mel_spec_latent_source} shows interesting results of generated audios, which remind us the conceptual view of the latent source in Figure \ref{fig:latentsource}.
For the given input track of Figure \ref{fig:mel_spec_latent_source} (a), we generate an audio track after masking all channels except for the fifth channel in the second head group, then the result sounds similar to the low-frequency band of drums (i.e., kick drum) as illustrated in Figure \ref{fig:mel_spec_latent_source} (b).
AMSS-Net can keep this channel and drop other drum-related channels to process ``apply lowpass to drums."
However, a latent source channel does not always contain a single class of instruments.
For example, the latent channel of the fourth row in the table deals with several instruments.
Some latent sources were not interpretable to the authors.
Generated samples are available online.\footnote{https://kuielab.github.io/AMSS-Net/latent\_source.html}

\subsection{Progressive Manipulation}
\label{sec:progresive}
\begin{figure}[t]
    \centering
    \includegraphics[width=\columnwidth]
    {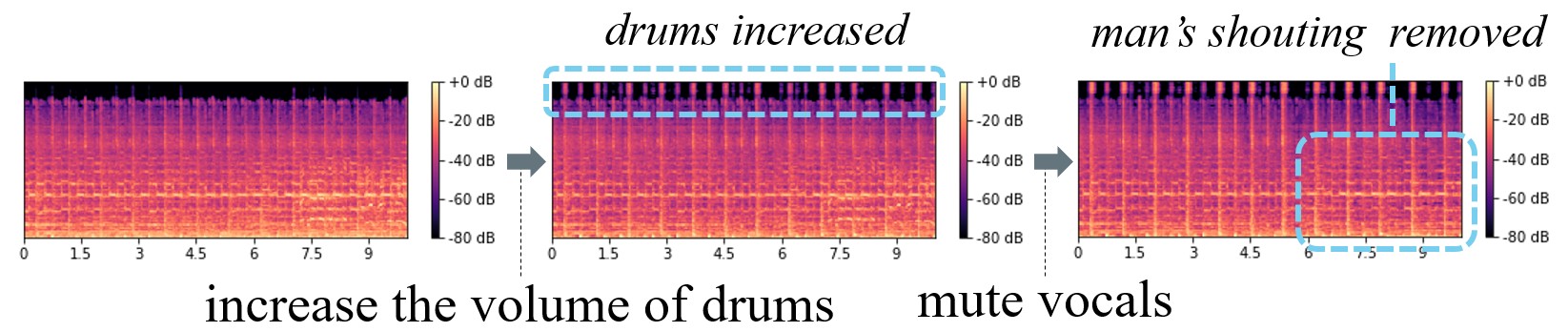}
    \caption{An Example of Progressive Manipulation}
    \label{fig:progressive}
\end{figure}

We show that we can repeatedly apply the proposed method to manipulated audio tracks, which is also known as Progressive Manipulation used in conversational systems described in  \cite{dwtc}.
Figure \ref{fig:progressive} shows an example of progressive manipulation. 

\begin{figure}[h]

\centering
\begin{tabular}{cc}
    \includegraphics[width=40mm]{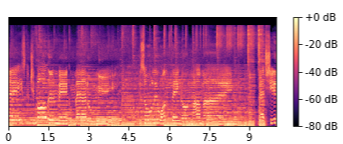} &
    \includegraphics[width=40mm]{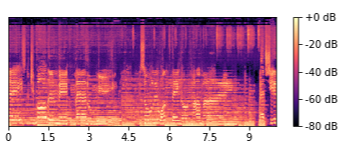}    \\
    (a) ground-truth target & (b) AMSS-Net \\
    \includegraphics[width=40mm]{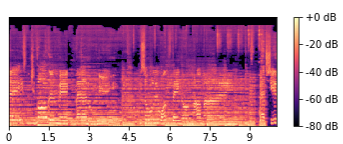} &
    \includegraphics[width=40mm]{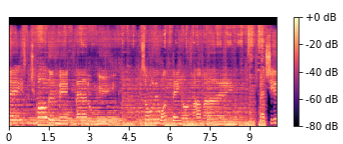}    \\
    (c) w/o CSA & (d) w/o SMPoCM \\
    
\end{tabular} 

\caption{Mel-Spectrogram Comparison after applying 20 times of `apply highpass to durms' in a progressive manner}
\label{fig:ablation}
\end{figure}

However, methods based on neural networks sometimes suffer from artifacts \cite{bss}, which are not present in the original source.
Although they sound negligible after a single manipulation task, they can be large enough to be perceived after progressively applying several.
To investigate artifacts created by progressive manipulation, we apply the same AMSS task ``apply highpass to drums'' to a track in a progressive manner. 
Figure (a) shows the Mel-spectrogram of the ground-truth target. 
We can observe blurred areas in the high-frequency range in Mel-spectrograms of Figure \ref{fig:ablation} (c) and (d). 
Compared to them, Figure \ref{fig:ablation} (b) is more similar to the ground-truth target.
Via hearing test, we observed perceivable artifacts \ref{fig:ablation} in the results of baselines.
Our AMSS-Net contains minor artifacts compare to them because each decoding block of AMSS-Net has a CSA mechanism, a unique structure that prevents unwanted noise generated by intermediate manipulated features.
Generated samples are available online\footnote{https://kuielab.github.io/AMSS-Net/progresive.html}.

\subsection{Controlling the level of audio effects}

\begin{figure}[h]

\centering
\begin{tabular}{cc}
    \includegraphics[width=40mm]{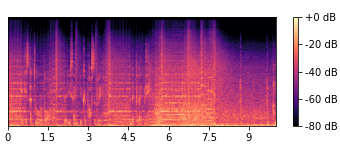} &
    \includegraphics[width=40mm]{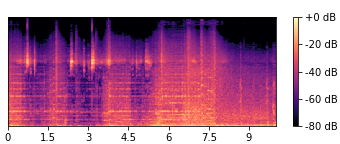}    \\
    (a) original & (b) heavy highpass to vocals \\
    \includegraphics[width=40mm]{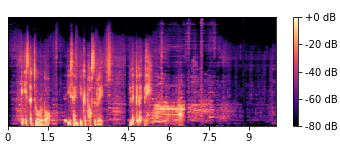} &
    \includegraphics[width=40mm]{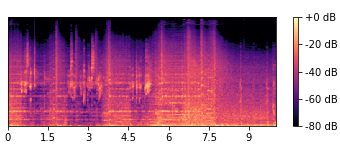}    \\
    (c) separate vocals & (d) medium highpass to vocals \\
    \includegraphics[width=40mm]{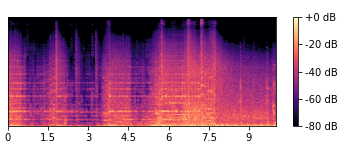} &
    \includegraphics[width=40mm]{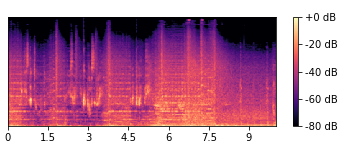}    \\
    (e) mute vocals & (f) light highpass to vocals \\
     
\end{tabular} 

\caption{Controlling the level of highpass with adjectives}
\label{fig:hml}
\end{figure}

As mentioned in \S \ref{sec:aml}, we can train AMSS-Net to perform more detailed AMSS such as ``apply heavy lowpass to vocals''.
As shown in Figure \ref{fig:hml}, users can control the level of audio effects by simply injecting adverbs instead of a laborious search for an appropriate parameter configuration.
Generated samples are available online\footnote{https://kuielab.github.io/AMSS-Net/control\_level.html}.

\subsection{Discussion}
\label{sec:discussion}
AMSS-Net shows promising results on several AMSS tasks. AMSS-Net can also be trained with a more complicated AMSS training dataset based on a realistic audio mixing dataset such as \textit{IDMT-SMT-Audio-Effects} dataset \cite{idmt}.
However, this study is limited to model relatively simpler AMSS tasks. One can extend our work to provide more complex AMSS tasks such as distortion and reverberation. Each AMSS task in this paper only deals with a single type of manipulations, but one can also extend our work to provide multiple types of tasks such as ``apply reverb to vocals and apply lowpass to drums'' at once.
Also, our work is easily extendable to support a more user-friendly interface. For example, adopting unsupervised training frameworks such as Mixture of Mixture (MoM) \cite{google} to train AMSS on annotated audio datasets such as clotho\cite{clotho} might enable a natural language query interface.

\section{Conclusion}

In this paper, we define a novel task called AMSS. 
We propose AMSS-Net, which generates feature maps in which each channel deals with a latent source and selectively manipulates them while preserving irrelevant features. AMSS-Net can perform several AMSS tasks, unlike previous models such as LaSAFT-GPoCM-Net.
The experimental results show that AMSS-Net outperforms baselines on several tasks.
Future work will extend it to provide more complex AMSS tasks such as distortion and reverberation by adopting state-of-the-art methods such as Generative Adversarial Networks (GAN).


\bibliographystyle{ACM-Reference-Format}
\bibliography{sample-base}


\begin{thebibliography}{38}


\ifx \showCODEN    \undefined \def \showCODEN     #1{\unskip}     \fi
\ifx \showDOI      \undefined \def \showDOI       #1{#1}\fi
\ifx \showISBNx    \undefined \def \showISBNx     #1{\unskip}     \fi
\ifx \showISBNxiii \undefined \def \showISBNxiii  #1{\unskip}     \fi
\ifx \showISSN     \undefined \def \showISSN      #1{\unskip}     \fi
\ifx \showLCCN     \undefined \def \showLCCN      #1{\unskip}     \fi
\ifx \shownote     \undefined \def \shownote      #1{#1}          \fi
\ifx \showarticletitle \undefined \def \showarticletitle #1{#1}   \fi
\ifx \showURL      \undefined \def \showURL       {\relax}        \fi
\providecommand\bibfield[2]{#2}
\providecommand\bibinfo[2]{#2}
\providecommand\natexlab[1]{#1}
\providecommand\showeprint[2][]{arXiv:#2}

\bibitem[\protect\citeauthoryear{Avendano}{Avendano}{2003}]%
        {algorithmic}
\bibfield{author}{\bibinfo{person}{Carlos Avendano}.}
  \bibinfo{year}{2003}\natexlab{}.
\newblock \showarticletitle{Frequency-domain source identification and
  manipulation in stereo mixes for enhancement, suppression and re-panning
  applications}. In \bibinfo{booktitle}{\emph{2003 IEEE Workshop on
  Applications of Signal Processing to Audio and Acoustics (IEEE Cat. No.
  03TH8684)}}. IEEE, \bibinfo{pages}{55--58}.
\newblock


\bibitem[\protect\citeauthoryear{Chen, Zhang, Xiao, Nie, Shao, Liu, and
  Chua}{Chen et~al\mbox{.}}{2017}]%
        {ca}
\bibfield{author}{\bibinfo{person}{Long Chen}, \bibinfo{person}{Hanwang Zhang},
  \bibinfo{person}{Jun Xiao}, \bibinfo{person}{Liqiang Nie},
  \bibinfo{person}{Jian Shao}, \bibinfo{person}{Wei Liu}, {and}
  \bibinfo{person}{Tat-Seng Chua}.} \bibinfo{year}{2017}\natexlab{}.
\newblock \showarticletitle{Sca-cnn: Spatial and channel-wise attention in
  convolutional networks for image captioning}. In
  \bibinfo{booktitle}{\emph{Proceedings of the IEEE conference on computer
  vision and pattern recognition}}. \bibinfo{pages}{5659--5667}.
\newblock


\bibitem[\protect\citeauthoryear{Choi, Kim, Chung, and Jung}{Choi
  et~al\mbox{.}}{2020a}]%
        {lasaft}
\bibfield{author}{\bibinfo{person}{Woosung Choi}, \bibinfo{person}{Minseok
  Kim}, \bibinfo{person}{Jaehwa Chung}, {and} \bibinfo{person}{Soonyoung
  Jung}.} \bibinfo{year}{2020}\natexlab{a}.
\newblock \showarticletitle{LaSAFT: Latent Source Attentive Frequency
  Transformation for Conditioned Source Separation}.
\newblock \bibinfo{journal}{\emph{arXiv preprint arXiv:2010.11631}}
  (\bibinfo{year}{2020}).
\newblock


\bibitem[\protect\citeauthoryear{Choi, Kim, Chung, Lee, and Jung}{Choi
  et~al\mbox{.}}{2020b}]%
        {tfctdf}
\bibfield{author}{\bibinfo{person}{Woosung Choi}, \bibinfo{person}{Minseok
  Kim}, \bibinfo{person}{Jaehwa Chung}, \bibinfo{person}{Daewon Lee}, {and}
  \bibinfo{person}{Soonyoung Jung}.} \bibinfo{year}{2020}\natexlab{b}.
\newblock \showarticletitle{Investigating u-nets with various intermediate
  blocks for spectrogram-based singing voice separation}. In
  \bibinfo{booktitle}{\emph{Proceedings of the 21th International Society for
  Music Information Retrieval Conference}}.
\newblock


\bibitem[\protect\citeauthoryear{{Chomsky}}{{Chomsky}}{1956}]%
        {cfg}
\bibfield{author}{\bibinfo{person}{N. {Chomsky}}.}
  \bibinfo{year}{1956}\natexlab{}.
\newblock \showarticletitle{Three models for the description of language}.
\newblock \bibinfo{journal}{\emph{IRE Transactions on Information Theory}}
  \bibinfo{volume}{2}, \bibinfo{number}{3} (\bibinfo{year}{1956}),
  \bibinfo{pages}{113--124}.
\newblock
\urldef\tempurl%
\url{https://doi.org/10.1109/TIT.1956.1056813}
\showDOI{\tempurl}


\bibitem[\protect\citeauthoryear{Drossos, Lipping, and Virtanen}{Drossos
  et~al\mbox{.}}{2020}]%
        {clotho}
\bibfield{author}{\bibinfo{person}{Konstantinos Drossos},
  \bibinfo{person}{Samuel Lipping}, {and} \bibinfo{person}{Tuomas Virtanen}.}
  \bibinfo{year}{2020}\natexlab{}.
\newblock \showarticletitle{Clotho: An audio captioning dataset}. In
  \bibinfo{booktitle}{\emph{ICASSP 2020-2020 IEEE International Conference on
  Acoustics, Speech and Signal Processing (ICASSP)}}. IEEE,
  \bibinfo{pages}{736--740}.
\newblock


\bibitem[\protect\citeauthoryear{Hochreiter and Schmidhuber}{Hochreiter and
  Schmidhuber}{1997}]%
        {lstm}
\bibfield{author}{\bibinfo{person}{Sepp Hochreiter} {and}
  \bibinfo{person}{Jürgen Schmidhuber}.} \bibinfo{year}{1997}\natexlab{}.
\newblock \showarticletitle{Long Short-Term Memory}.
\newblock \bibinfo{journal}{\emph{Neural Computation}} \bibinfo{volume}{9},
  \bibinfo{number}{8} (\bibinfo{year}{1997}), \bibinfo{pages}{1735--1780}.
\newblock


\bibitem[\protect\citeauthoryear{Jansson, Humphrey, Montecchio, Bittner, Kumar,
  and Weyde}{Jansson et~al\mbox{.}}{2017}]%
        {svs_unet}
\bibfield{author}{\bibinfo{person}{Andreas Jansson}, \bibinfo{person}{Eric
  Humphrey}, \bibinfo{person}{Nicola Montecchio}, \bibinfo{person}{Rachel
  Bittner}, \bibinfo{person}{Aparna Kumar}, {and} \bibinfo{person}{Tillman
  Weyde}.} \bibinfo{year}{2017}\natexlab{}.
\newblock \showarticletitle{Singing voice separation with deep u-net
  convolutional networks}. In \bibinfo{booktitle}{\emph{18th International
  Society for Music Information Retrieval Conference}}.
  \bibinfo{pages}{745--751}.
\newblock


\bibitem[\protect\citeauthoryear{Kingma and Ba}{Kingma and Ba}{2015}]%
        {adam}
\bibfield{author}{\bibinfo{person}{Diederik~P. Kingma} {and}
  \bibinfo{person}{Jimmy Ba}.} \bibinfo{year}{2015}\natexlab{}.
\newblock \showarticletitle{Adam: {A} Method for Stochastic Optimization}. In
  \bibinfo{booktitle}{\emph{3rd International Conference on Learning
  Representations, {ICLR} 2015, San Diego, CA, USA, May 7-9, 2015, Conference
  Track Proceedings}}, \bibfield{editor}{\bibinfo{person}{Yoshua Bengio} {and}
  \bibinfo{person}{Yann LeCun}} (Eds.).
\newblock
\urldef\tempurl%
\url{http://arxiv.org/abs/1412.6980}
\showURL{%
\tempurl}


\bibitem[\protect\citeauthoryear{Li, Qi, Lukasiewicz, and Torr}{Li
  et~al\mbox{.}}{2020}]%
        {manigan}
\bibfield{author}{\bibinfo{person}{Bowen Li}, \bibinfo{person}{Xiaojuan Qi},
  \bibinfo{person}{Thomas Lukasiewicz}, {and} \bibinfo{person}{Philip~HS
  Torr}.} \bibinfo{year}{2020}\natexlab{}.
\newblock \showarticletitle{Manigan: Text-guided image manipulation}. In
  \bibinfo{booktitle}{\emph{Proceedings of the IEEE/CVF Conference on Computer
  Vision and Pattern Recognition}}. \bibinfo{pages}{7880--7889}.
\newblock


\bibitem[\protect\citeauthoryear{Liu and Yang}{Liu and Yang}{2019}]%
        {dilatedlstm}
\bibfield{author}{\bibinfo{person}{Jen-Yu Liu} {and} \bibinfo{person}{Yi-Hsuan
  Yang}.} \bibinfo{year}{2019}\natexlab{}.
\newblock \showarticletitle{Dilated Convolution with Dilated GRU for Music
  Source Separation}. In \bibinfo{booktitle}{\emph{Proceedings of the
  Twenty-Eighth International Joint Conference on Artificial Intelligence,
  {IJCAI-19}}}. \bibinfo{publisher}{International Joint Conferences on
  Artificial Intelligence Organization}, \bibinfo{pages}{4718--4724}.
\newblock
\urldef\tempurl%
\url{https://doi.org/10.24963/ijcai.2019/655}
\showDOI{\tempurl}


\bibitem[\protect\citeauthoryear{Liu, De~Nadai, Cai, Li, Alameda-Pineda, Sebe,
  and Lepri}{Liu et~al\mbox{.}}{2020}]%
        {dwtc}
\bibfield{author}{\bibinfo{person}{Yahui Liu}, \bibinfo{person}{Marco
  De~Nadai}, \bibinfo{person}{Deng Cai}, \bibinfo{person}{Huayang Li},
  \bibinfo{person}{Xavier Alameda-Pineda}, \bibinfo{person}{Nicu Sebe}, {and}
  \bibinfo{person}{Bruno Lepri}.} \bibinfo{year}{2020}\natexlab{}.
\newblock \showarticletitle{Describe What to Change: A Text-guided Unsupervised
  Image-to-Image Translation Approach}. In
  \bibinfo{booktitle}{\emph{Proceedings of the 28th ACM International
  Conference on Multimedia}}. \bibinfo{pages}{1357--1365}.
\newblock


\bibitem[\protect\citeauthoryear{Mart{\'\i}nez~Ram{\'\i}rez, Benetos, and
  Reiss}{Mart{\'\i}nez~Ram{\'\i}rez et~al\mbox{.}}{2020}]%
        {marco2}
\bibfield{author}{\bibinfo{person}{Marco~A Mart{\'\i}nez~Ram{\'\i}rez},
  \bibinfo{person}{Emmanouil Benetos}, {and} \bibinfo{person}{Joshua~D Reiss}.}
  \bibinfo{year}{2020}\natexlab{}.
\newblock \showarticletitle{Deep learning for black-box modeling of audio
  effects}.
\newblock \bibinfo{journal}{\emph{APPLIED SCIENCES-BASEL}}
  \bibinfo{volume}{10}, \bibinfo{number}{2} (\bibinfo{year}{2020}).
\newblock


\bibitem[\protect\citeauthoryear{Mart{\'\i}nez~Ram{\'\i}rez and
  Reiss}{Mart{\'\i}nez~Ram{\'\i}rez and Reiss}{2019}]%
        {marco1}
\bibfield{author}{\bibinfo{person}{Marco~A Mart{\'\i}nez~Ram{\'\i}rez} {and}
  \bibinfo{person}{Joshua~D Reiss}.} \bibinfo{year}{2019}\natexlab{}.
\newblock \showarticletitle{Modeling nonlinear audio effects with end-to-end
  deep neural networks}. In \bibinfo{booktitle}{\emph{ICASSP 2019-2019 IEEE
  International Conference on Acoustics, Speech and Signal Processing
  (ICASSP)}}. IEEE, \bibinfo{pages}{171--175}.
\newblock


\bibitem[\protect\citeauthoryear{Mart{\'\i}nez~Ram{\'\i}rez, Stoller, and
  Moffat}{Mart{\'\i}nez~Ram{\'\i}rez et~al\mbox{.}}{2021a}]%
        {marco3}
\bibfield{author}{\bibinfo{person}{Marco~A Mart{\'\i}nez~Ram{\'\i}rez},
  \bibinfo{person}{Daniel Stoller}, {and} \bibinfo{person}{David Moffat}.}
  \bibinfo{year}{2021}\natexlab{a}.
\newblock \showarticletitle{A Deep Learning Approach to Intelligent Drum Mixing
  with the Wave-U-Net}.
\newblock \bibinfo{journal}{\emph{Journal of the Audio Engineering Society}}
  \bibinfo{volume}{69}, \bibinfo{number}{3} (\bibinfo{year}{2021}),
  \bibinfo{pages}{142--151}.
\newblock


\bibitem[\protect\citeauthoryear{Mart{\'\i}nez~Ram{\'\i}rez, Stoller, and
  Moffat}{Mart{\'\i}nez~Ram{\'\i}rez et~al\mbox{.}}{2021b}]%
        {marco4}
\bibfield{author}{\bibinfo{person}{Marco~A Mart{\'\i}nez~Ram{\'\i}rez},
  \bibinfo{person}{Daniel Stoller}, {and} \bibinfo{person}{David Moffat}.}
  \bibinfo{year}{2021}\natexlab{b}.
\newblock \showarticletitle{A Deep Learning Approach to Intelligent Drum Mixing
  with the Wave-U-Net}.
\newblock \bibinfo{journal}{\emph{Journal of the Audio Engineering Society}}
  \bibinfo{volume}{69}, \bibinfo{number}{3} (\bibinfo{year}{2021}),
  \bibinfo{pages}{142--151}.
\newblock


\bibitem[\protect\citeauthoryear{Matz, Cano, and Abe{\ss}er}{Matz
  et~al\mbox{.}}{2015}]%
        {jazz1}
\bibfield{author}{\bibinfo{person}{D. Matz}, \bibinfo{person}{Estefan{\'i}a
  Cano}, {and} \bibinfo{person}{J. Abe{\ss}er}.}
  \bibinfo{year}{2015}\natexlab{}.
\newblock \showarticletitle{New Sonorities for Early Jazz Recordings Using
  Sound Source Separation and Automatic Mixing Tools}. In
  \bibinfo{booktitle}{\emph{ISMIR}}.
\newblock


\bibitem[\protect\citeauthoryear{Meseguer-Brocal and Peeters}{Meseguer-Brocal
  and Peeters}{2019}]%
        {cunet}
\bibfield{author}{\bibinfo{person}{Gabriel Meseguer-Brocal} {and}
  \bibinfo{person}{Geoffroy Peeters}.} \bibinfo{year}{2019}\natexlab{}.
\newblock \showarticletitle{CONDITIONED-U-NET: Introducing a Control Mechanism
  in the U-net For Multiple Source Separations.}. In
  \bibinfo{booktitle}{\emph{20th International Society for Music Information
  Retrieval Conference}}, \bibfield{editor}{\bibinfo{person}{ISMIR}} (Ed.).
\newblock


\bibitem[\protect\citeauthoryear{Mimilakis, Cano, Abe{\ss}er, and
  Schuller}{Mimilakis et~al\mbox{.}}{2016}]%
        {jazz2}
\bibfield{author}{\bibinfo{person}{Stylianos~Ioannis Mimilakis},
  \bibinfo{person}{Estefan{\i}a Cano}, \bibinfo{person}{Jakob Abe{\ss}er},
  {and} \bibinfo{person}{Gerald Schuller}.} \bibinfo{year}{2016}\natexlab{}.
\newblock \showarticletitle{New sonorities for jazz recordings: Separation and
  mixing using deep neural networks}. In \bibinfo{booktitle}{\emph{2nd AES
  Workshop on Intelligent Music Production}}, Vol.~\bibinfo{volume}{13}.
\newblock


\bibitem[\protect\citeauthoryear{Pennington, Socher, and Manning}{Pennington
  et~al\mbox{.}}{2014}]%
        {glove}
\bibfield{author}{\bibinfo{person}{Jeffrey Pennington},
  \bibinfo{person}{Richard Socher}, {and} \bibinfo{person}{Christopher~D
  Manning}.} \bibinfo{year}{2014}\natexlab{}.
\newblock \showarticletitle{Glove: Global Vectors for Word Representation.}. In
  \bibinfo{booktitle}{\emph{EMNLP}}, Vol.~\bibinfo{volume}{14}.
  \bibinfo{pages}{1532--1543}.
\newblock


\bibitem[\protect\citeauthoryear{Perez, Strub, de~Vries, Dumoulin, and
  Courville}{Perez et~al\mbox{.}}{2018}]%
        {film}
\bibfield{author}{\bibinfo{person}{Ethan Perez}, \bibinfo{person}{Florian
  Strub}, \bibinfo{person}{Harm de Vries}, \bibinfo{person}{Vincent Dumoulin},
  {and} \bibinfo{person}{Aaron~C Courville}.} \bibinfo{year}{2018}\natexlab{}.
\newblock \showarticletitle{FiLM: Visual Reasoning with a General Conditioning
  Layer}. In \bibinfo{booktitle}{\emph{AAAI}}.
\newblock


\bibitem[\protect\citeauthoryear{Rafii, Liutkus, St{\"o}ter, Mimilakis, and
  Bittner}{Rafii et~al\mbox{.}}{2017}]%
        {musdb18}
\bibfield{author}{\bibinfo{person}{Zafar Rafii}, \bibinfo{person}{Antoine
  Liutkus}, \bibinfo{person}{Fabian-Robert St{\"o}ter},
  \bibinfo{person}{Stylianos~Ioannis Mimilakis}, {and} \bibinfo{person}{Rachel
  Bittner}.} \bibinfo{year}{2017}\natexlab{}.
\newblock \bibinfo{title}{{MUSDB18 - a corpus for music separation}}.
\newblock
\newblock
\urldef\tempurl%
\url{https://doi.org/10.5281/zenodo.1117371}
\showDOI{\tempurl}
\newblock
\shownote{MUSDB18: a corpus for music source separation.}


\bibitem[\protect\citeauthoryear{Ram{\'\i}rez and Reiss}{Ram{\'\i}rez and
  Reiss}{2018}]%
        {marco0}
\bibfield{author}{\bibinfo{person}{Mart{\'\i}nez Ram{\'\i}rez} {and}
  \bibinfo{person}{Joshua~D Reiss}.} \bibinfo{year}{2018}\natexlab{}.
\newblock \showarticletitle{End-to-end equalization with convolutional neural
  networks}. In \bibinfo{booktitle}{\emph{21st International Conference on
  Digital Audio Effects (DAFx-18)}}.
\newblock


\bibitem[\protect\citeauthoryear{Samuel, Ganeshan, and Naradowsky}{Samuel
  et~al\mbox{.}}{2020}]%
        {meta}
\bibfield{author}{\bibinfo{person}{David Samuel}, \bibinfo{person}{Aditya
  Ganeshan}, {and} \bibinfo{person}{Jason Naradowsky}.}
  \bibinfo{year}{2020}\natexlab{}.
\newblock \showarticletitle{Meta-learning Extractors for Music Source
  Separation}. In \bibinfo{booktitle}{\emph{ICASSP 2020-2020 IEEE International
  Conference on Acoustics, Speech and Signal Processing (ICASSP)}}. IEEE,
  \bibinfo{pages}{816--820}.
\newblock


\bibitem[\protect\citeauthoryear{Schuster and Paliwal}{Schuster and
  Paliwal}{1997}]%
        {birnn}
\bibfield{author}{\bibinfo{person}{Mike Schuster} {and}
  \bibinfo{person}{Kuldip~K Paliwal}.} \bibinfo{year}{1997}\natexlab{}.
\newblock \showarticletitle{Bidirectional recurrent neural networks}.
\newblock \bibinfo{journal}{\emph{IEEE transactions on Signal Processing}}
  \bibinfo{volume}{45}, \bibinfo{number}{11} (\bibinfo{year}{1997}),
  \bibinfo{pages}{2673--2681}.
\newblock


\bibitem[\protect\citeauthoryear{Stein, Abe{\ss}er, Dittmar, and
  Schuller}{Stein et~al\mbox{.}}{2010}]%
        {idmt}
\bibfield{author}{\bibinfo{person}{Michael Stein}, \bibinfo{person}{Jakob
  Abe{\ss}er}, \bibinfo{person}{Christian Dittmar}, {and}
  \bibinfo{person}{Gerald Schuller}.} \bibinfo{year}{2010}\natexlab{}.
\newblock \showarticletitle{Automatic detection of audio effects in guitar and
  bass recordings}. In \bibinfo{booktitle}{\emph{Audio Engineering Society
  Convention 128}}. Audio Engineering Society.
\newblock


\bibitem[\protect\citeauthoryear{Steinmetz, Pons, Pascual, and
  Serr{\`a}}{Steinmetz et~al\mbox{.}}{2020}]%
        {christian1}
\bibfield{author}{\bibinfo{person}{Christian~J Steinmetz},
  \bibinfo{person}{Jordi Pons}, \bibinfo{person}{Santiago Pascual}, {and}
  \bibinfo{person}{Joan Serr{\`a}}.} \bibinfo{year}{2020}\natexlab{}.
\newblock \showarticletitle{Automatic multitrack mixing with a differentiable
  mixing console of neural audio effects}.
\newblock \bibinfo{journal}{\emph{arXiv preprint arXiv:2010.10291}}
  (\bibinfo{year}{2020}).
\newblock


\bibitem[\protect\citeauthoryear{Steinmetz and Reiss}{Steinmetz and
  Reiss}{2021}]%
        {christian}
\bibfield{author}{\bibinfo{person}{Christian~J Steinmetz} {and}
  \bibinfo{person}{Joshua~D Reiss}.} \bibinfo{year}{2021}\natexlab{}.
\newblock \showarticletitle{Efficient Neural Networks for Real-time Analog
  Audio Effect Modeling}.
\newblock \bibinfo{journal}{\emph{arXiv preprint arXiv:2102.06200}}
  (\bibinfo{year}{2021}).
\newblock


\bibitem[\protect\citeauthoryear{Uhlich, Porcu, Giron, Enenkl, Kemp, Takahashi,
  and Mitsufuji}{Uhlich et~al\mbox{.}}{2017}]%
        {blend}
\bibfield{author}{\bibinfo{person}{Stefan Uhlich}, \bibinfo{person}{Marcello
  Porcu}, \bibinfo{person}{Franck Giron}, \bibinfo{person}{Michael Enenkl},
  \bibinfo{person}{Thomas Kemp}, \bibinfo{person}{Naoya Takahashi}, {and}
  \bibinfo{person}{Yuki Mitsufuji}.} \bibinfo{year}{2017}\natexlab{}.
\newblock \showarticletitle{Improving music source separation based on deep
  neural networks through data augmentation and network blending}. In
  \bibinfo{booktitle}{\emph{2017 IEEE International Conference on Acoustics,
  Speech and Signal Processing (ICASSP)}}. IEEE, \bibinfo{pages}{261--265}.
\newblock


\bibitem[\protect\citeauthoryear{Vaswani, Shazeer, Parmar, Uszkoreit, Jones,
  Gomez, Kaiser, and Polosukhin}{Vaswani et~al\mbox{.}}{2017}]%
        {transformer}
\bibfield{author}{\bibinfo{person}{Ashish Vaswani}, \bibinfo{person}{Noam
  Shazeer}, \bibinfo{person}{Niki Parmar}, \bibinfo{person}{Jakob Uszkoreit},
  \bibinfo{person}{Llion Jones}, \bibinfo{person}{Aidan~N Gomez},
  \bibinfo{person}{{\L}ukasz Kaiser}, {and} \bibinfo{person}{Illia
  Polosukhin}.} \bibinfo{year}{2017}\natexlab{}.
\newblock \showarticletitle{Attention is all you need}. In
  \bibinfo{booktitle}{\emph{Advances in neural information processing
  systems}}. \bibinfo{pages}{5998--6008}.
\newblock


\bibitem[\protect\citeauthoryear{Vincent, Gribonval, and F{\'e}votte}{Vincent
  et~al\mbox{.}}{2006}]%
        {bss}
\bibfield{author}{\bibinfo{person}{Emmanuel Vincent}, \bibinfo{person}{R{\'e}mi
  Gribonval}, {and} \bibinfo{person}{C{\'e}dric F{\'e}votte}.}
  \bibinfo{year}{2006}\natexlab{}.
\newblock \showarticletitle{Performance measurement in blind audio source
  separation}.
\newblock \bibinfo{journal}{\emph{IEEE transactions on audio, speech, and
  language processing}} \bibinfo{volume}{14}, \bibinfo{number}{4}
  (\bibinfo{year}{2006}), \bibinfo{pages}{1462--1469}.
\newblock


\bibitem[\protect\citeauthoryear{Wisdom, Tzinis, Erdogan, Weiss, Wilson, and
  Hershey}{Wisdom et~al\mbox{.}}{2020}]%
        {google}
\bibfield{author}{\bibinfo{person}{Scott Wisdom}, \bibinfo{person}{Efthymios
  Tzinis}, \bibinfo{person}{Hakan Erdogan}, \bibinfo{person}{Ron~J. Weiss},
  \bibinfo{person}{Kevin Wilson}, {and} \bibinfo{person}{John~R. Hershey}.}
  \bibinfo{year}{2020}\natexlab{}.
\newblock \showarticletitle{Unsupervised Sound Separation Using Mixture
  Invariant Training}. In \bibinfo{booktitle}{\emph{NeurIPS}}.
\newblock
\urldef\tempurl%
\url{https://arxiv.org/pdf/2006.12701.pdf}
\showURL{%
\tempurl}


\bibitem[\protect\citeauthoryear{Wright, Damskägg, Juvela, and
  Välimäki}{Wright et~al\mbox{.}}{2020}]%
        {amp}
\bibfield{author}{\bibinfo{person}{Alec Wright}, \bibinfo{person}{Eero-Pekka
  Damskägg}, \bibinfo{person}{Lauri Juvela}, {and} \bibinfo{person}{Vesa
  Välimäki}.} \bibinfo{year}{2020}\natexlab{}.
\newblock \showarticletitle{Real-Time Guitar Amplifier Emulation with Deep
  Learning}.
\newblock \bibinfo{journal}{\emph{Applied Sciences}} \bibinfo{volume}{10},
  \bibinfo{number}{3} (\bibinfo{year}{2020}).
\newblock
\showISSN{2076-3417}
\urldef\tempurl%
\url{https://doi.org/10.3390/app10030766}
\showDOI{\tempurl}


\bibitem[\protect\citeauthoryear{Wyse}{Wyse}{2017}]%
        {trans}
\bibfield{author}{\bibinfo{person}{Lonce Wyse}.}
  \bibinfo{year}{2017}\natexlab{}.
\newblock \showarticletitle{Audio spectrogram representations for processing
  with convolutional neural networks}.
\newblock \bibinfo{journal}{\emph{arXiv preprint arXiv:1706.09559}}
  (\bibinfo{year}{2017}).
\newblock


\bibitem[\protect\citeauthoryear{Yin, Luo, Xiong, and Zeng}{Yin
  et~al\mbox{.}}{2019}]%
        {phasen}
\bibfield{author}{\bibinfo{person}{Dacheng Yin}, \bibinfo{person}{Chong Luo},
  \bibinfo{person}{Zhiwei Xiong}, {and} \bibinfo{person}{Wenjun Zeng}.}
  \bibinfo{year}{2019}\natexlab{}.
\newblock \showarticletitle{PHASEN: A Phase-and-Harmonics-Aware Speech
  Enhancement Network}.
\newblock \bibinfo{journal}{\emph{arXiv preprint arXiv:1911.04697}}
  (\bibinfo{year}{2019}).
\newblock


\bibitem[\protect\citeauthoryear{Yu, Lin, Yang, Shen, Lu, and Huang}{Yu
  et~al\mbox{.}}{2018}]%
        {inpainting}
\bibfield{author}{\bibinfo{person}{Jiahui Yu}, \bibinfo{person}{Zhe Lin},
  \bibinfo{person}{Jimei Yang}, \bibinfo{person}{Xiaohui Shen},
  \bibinfo{person}{Xin Lu}, {and} \bibinfo{person}{Thomas~S Huang}.}
  \bibinfo{year}{2018}\natexlab{}.
\newblock \showarticletitle{Generative image inpainting with contextual
  attention}. In \bibinfo{booktitle}{\emph{Proceedings of the IEEE conference
  on computer vision and pattern recognition}}. \bibinfo{pages}{5505--5514}.
\newblock


\bibitem[\protect\citeauthoryear{Yuan, Wang, Li, Unoki, and Wang}{Yuan
  et~al\mbox{.}}{2019}]%
        {sa}
\bibfield{author}{\bibinfo{person}{Weitao Yuan}, \bibinfo{person}{Shengbei
  Wang}, \bibinfo{person}{Xiangrui Li}, \bibinfo{person}{Masashi Unoki}, {and}
  \bibinfo{person}{Wenwu Wang}.} \bibinfo{year}{2019}\natexlab{}.
\newblock \showarticletitle{A Skip Attention Mechanism for Monaural Singing
  Voice Separation}.
\newblock \bibinfo{journal}{\emph{IEEE Signal Processing Letters}}
  \bibinfo{volume}{26}, \bibinfo{number}{10} (\bibinfo{year}{2019}),
  \bibinfo{pages}{1481--1485}.
\newblock


\bibitem[\protect\citeauthoryear{Zhu, Park, Isola, and Efros}{Zhu
  et~al\mbox{.}}{2017}]%
        {cyclegan}
\bibfield{author}{\bibinfo{person}{Jun-Yan Zhu}, \bibinfo{person}{Taesung
  Park}, \bibinfo{person}{Phillip Isola}, {and} \bibinfo{person}{Alexei~A
  Efros}.} \bibinfo{year}{2017}\natexlab{}.
\newblock \showarticletitle{Unpaired image-to-image translation using
  cycle-consistent adversarial networks}. In
  \bibinfo{booktitle}{\emph{Proceedings of the IEEE international conference on
  computer vision}}. \bibinfo{pages}{2223--2232}.
\newblock


\end{thebibliography}


\end{document}